\documentclass[showpacs, twocolumn, prl]{revtex4}
\usepackage{graphicx}
\usepackage{times}

\begin{document}

\title{
Quantum Monte Carlo study of MnO solid
      }

\author{
Ji-Woo Lee
}
\affiliation{Department of Physics, Duke University, Durham, NC 27708-0305}
\author{Lubos Mitas and Lucas K. Wagner}
\affiliation{
Center for High Performance Simulation and
Department of Physics, North Carolina State University,
 Raleigh, NC 27695-8202}

\date{\today}

\begin{abstract}
 Electronic structure of the manganese oxide solid is studied by the
 quantum Monte Carlo (QMC) methods. 
 The trial wavefunctions are built using orbitals
 from unrestricted Hartree-Fock and Density Functional 
 Theory, and the electron-electron correlation is recovered by
 the fixed-node QMC.
 The correlation effects are significant and QMC
 estimations of the gap and cohesion show a very good agreement with 
 experiment. Comparison with hybrid functional results points
 out the importance of the exact exchange for improvement of the
  Density Functional description of transition metal oxide systems.  

\end{abstract}

\maketitle

 Transition metal compounds and transition metal oxides~(TMOs) 
 in particular belong to the most complex and important types
 of solid materials. TMOs exhibit a multitude of collective 
 effects such as  ferro-, ferri- and anti-ferromagnetism,
  ferroelectricity,
 superconductivity, etc {\cite{Brandowold, Brandownew}}. 
The electronic structure
 of these systems poses a real challenge both for theory and  experiment
 and TMOs have remained on a forefront of condensed matter research for
 decades  
~{\cite{Zaanen,Pask,Oguchi,Terakura,Svane,Szotek,Norman,Anisimov,Hugel}}.  
Among the TMO solids,
MnO and NiO have become paradigmatic examples of
 strong electron-electron correlation effects
 and antiferromagnetic ordering.
 Both MnO and NiO should nominally exhibit partially filled
 levels which would lead to a metallic ground state, however, experiments 
 revealed that these systems are actually antiferromagnetic 
 insulators with large gaps. 
The qualitative picture of electronic structure is explained either
  by a Mott-Hubbard
 mechanism implying that the gap results from a large local
 Coulomb repulsion in doubly occupied $d$-states, or by a charge
 transfer from transition metal $d$-states to oxygen $p$ levels with the gap having
 $p\to  d$ like character or a by a combined regime in between
 ~\cite{Zaanen}.
 The localized $d$ subshells exhibit unpaired spins which at low temperatures  
 order into an antiferromagnetic AF II insulator with alternating
 spin (111) planes in cubic rocksalt structure.

 The electronic structure of MnO and NiO have been studied by a number 
 of theoretical approaches, most notably by the spin-polarized 
 Density Functional Theory (DFT) with several types of functionals.  
 Augmented spherical wave local density approximation calculations~\cite{Oguchi} and
 subsequent works~\cite{Terakura} explained the stability of
 the AF II ordering in NiO and MnO. Very recently, the rhombohedral distortion of 
 MnO was successfully described by accurate DFT calculations {\cite{Pask}}. However,
 the commonly used DFT functionals 
are less reliable for predicting other key properties
 of strongly
 correlated systems. For example, DFT band gaps can be underestimated by
 a factor of 3-10 or even absent, leading thus to false
 metallic states. Therefore
 a variety of DFT modifications  
 such as
 self-interaction correction~\cite{Svane, Szotek},
 orbital-polarization corrections~\cite{Norman},
 on-site Coulomb interaction (LDA+U approach)~\cite{Anisimov, Hugel}
 and others,
  have been suggested to improve the description 
 of gaps and other electronic properties.
 On the other hand, 
 Towler {\it et al.}~\cite{Towler}  studied MnO and NiO with 
 unrestricted Hartree-Fock theory (UHF).
 UHF is seemingly a poor method since it neglects
 the electron correlation completely. Nevertheless,
 for transition elements the exchange, which is exact in the HF theory, 
 is at least as important as correlation, especially for metallic
 ions with an effective $d-$subshells occupation
 close to half-filling.
 The UHF results confirmed the crucial role of
 exchange in TMOs and provided a complementary picture to DFT
 with overestimated gaps and underestimated cohesion, but 
 also with the correct AF order, magnetic moments within 10\% from experiments
 and reasonably accurate lattice constants.
 Moreover, unlike DFT approaches which predict
 insulator only for the AF II ground state, UHF keeps the gap open
 also for the ferromagnetic or any spin-disordered phases. 
 This agrees with experiment which shows that MnO is 
  an insulator well above the
 N\'{e}el temperature $T_N \sim 118 {\rm K}$ since
 the spin ordering Mn-O-Mn superexchange mechanism is very weak and 
  hardly affected by spin flips on localized ions.

 In this letter,
 we take a fresh look on the MnO system in the framework of
 many-body quantum Monte Carlo method (for example, Ref. \cite{MitasRMP}
and references therein).  Our
 aim is to understand and quantify the impact of  
  explicit treatment of both exact exchange and correlation
 on the key properties such as cohesion and band gap.
In $sp$ systems 
 the QMC method was very successful
 in capturing electron correlation effects  
 for a large number of valence electrons 
  \cite{MitasRMP}.
It is therefore both important and interesting
 to test the QMC performance on
 challenging problems such as TMOs, extending  
the previous attempts to apply QMC to
 NiO system \cite{Tanaka,Needs}.

 We employ variational Monte Carlo and the diffusion Monte Carlo
 (DMC) methods. The trial/variational wavefunction 
is expressed as a product
 of Slater determinants of single-particle spin-up and spin-down
 orbitals ($\{\varphi_{\alpha} \},  \{ \varphi_{\beta}\} $) 
multiplied by a Jastrow correlation factor {\cite{Schmidt,Umrigar}},
\begin{equation}
\Psi_T = {\rm Det} \{ \varphi_{\alpha}\}
{\rm Det} \{ \varphi_{\beta}\}
\exp \left[ \sum_{I,i<j} u(r_{iI},r_{jI},r_{ij}) \right] \; \; ,
\end{equation}
where $I$ corresponds to the ions, $i,j$ to the electrons and
$r_{iI},r_{jI},r_{ij}$ to the distances.
 Similarly to the  previous work {\cite{Lubos3,Tanaka, MitasRMP}}
 the correlation function consists of a linear combination
 of electron-electron and electron-ion terms and involves  9 variational
 parameters.
 The DMC method is used to remove the major part
 of variational bias which is inherent to the variational methods.
 DMC is based on the property that the projection
 $\lim_{\tau\to \infty}\exp(-\tau H)\Psi_T$ is proportional
 to the ground state for an arbitrary $\Psi_T$
 with the same symmetry and non-zero overlap \cite{Lubos1}.
 The fermion sign problem caused by
 the antisymmetry of electrons  is avoided by
 the commonly used fixed-node approximation. 

In our calculations the core electrons are eliminated by pseudopotentials. 
For transition elements
the choice of accurate
 pseudopotentials is far from trivial.
In many TMO pseudopotential calculations
 with plane
 wave basis sets only the outermost $d$ and $s$ states 
 are included into the valence space.
  However,  earlier work {\cite {LubosFe}}
 indicated that the errors from elimination of the so-called
 semicore states, such as  
  $3s$ and $3p$ in the $3d$ transition series, could be significant.
 To elucidate this aspect,  Table I. provides a
 comparison of the $s \to d$ excitation energies for the Mn atom
 as obtained by the DMC method using scalar relativistic
 large-core (Ar) {\cite {Pacios}}
 and small-core (Ne) pseudopotentials {\cite {Dolg}}.
 It is evident that inclusion of $3s$ and $3p$ states
 into the valence space is important. 
 The large-core (Ar) pseudopotentials show errors of 
 the order $\approx$ 0.5-0.6 eV
 while the  Ne-core  pseudopotentials are much more accurate
 with typical errors $\approx$ 0.1 eV, what is comparable to the
 bias from
 the fixed-node approximation {\cite {MitasRMP}}.
 The physical reason of such behavior is well-known and 
 stems from the
 spatial distribution of the $3s$, $3p$ electrons 
 which is similar to the one of
 $3d$ electrons since they occupy the same principal shell.
 The price which one has to pay for using small core
 is, of course, significant. The magnitude of total energy,
 which is one of the key measures of QMC computational demands,
 increases roughly eightfold. It is also useful to notice that
 the DFT results 
 for these excitations differ from experiments by
 0.6-1 eV, clearly indicating the large contributions from exchange
 and correlation (Tab. I.).

\begin{table}[ht]
\vspace{0.05cm}
\caption{
 The excitation energies [eV] $s\to d$  of the Mn atom as calculated by
 UHF, DFT (BLYP and B3LYP)   
 and DMC methods.
 We used the Ne-core scalar relativistic pseudopotentials except
for the DMC(Ar) calculations which employed the Ar-core pseudopotentials. 
In the last
column are the experimental energies. Our DFT/BLYP results are close to similar
 calculations done previously, see Ref. \cite{Martin}. 
} 
\begin{center}
\begin{tabular}{l c c c c c c}
 & {UHF} & {BLYP} & {B3LYP} & DMC(Ar) & DMC(Ne) & {Exp.} \\
\hline
$s\to d$ & 3.5  & 1.2  & 1.6  & 1.6(1) &  2.2(1) & 2.1 \\
$s^2\to d^2$ & 9.1  & 4.4  & 5.2  & 5.2(1) &  5.8(1) & 5.6 \\
\hline
\end{tabular}
\end{center}
\label{tab_xx}
\end{table}

For the MnO solid we first carried out calculations 
 with the spin-unrestricted Hartree-Fock and
 DFT (B3LYP and PW86) methods and Ne-core pseudopotentials
using the CRYSTAL98/03 packages {\cite {CRYSTAL98}}.
 The orbitals were expanded in gaussian basis sets with
 $(12s,12p,7d)$ gaussians contracted to
$[3s,3p,2d]$ 
and $(8s,8p,1d)$ contracted to $[4s,4p,1d]$ for Mn and O atoms, respectively. 
 Figure 1 shows the band structure of MnO solid which is obtained from
UHF~(a), B3LYP~(b), PW86~(c) methods.
Note that B3LYP hybrid functional contains  20\% of the 
Hartree-Fock exchange 
so that it ``interpolates" between the exact HF exchange
and the effective local DFT exchange limits {\cite{Becke}} and
it often provides 
 an improved picture of excitations both in molecules and solids.

 In QMC the MnO solid is represented  by a supercell with periodic
 boundary conditions. This way of simulating the actual solid
 involves finite size errors which scale as $1/N$ where $N$ is the number of
 atoms in the supercell \cite{ Ceperley, Kent}.
 The finite size errors affect mainly the 
 estimation of cohesive energy where one needs to calculate the energy per
 primitive cell vs. isolated atoms.  
 In order to filter out the finite size bias we have
 carried out VMC calculations of supercells 
  with 8, 12, 16, 20 and 24 atom/supercell.
The most accurate and extensive fixed-node DMC calculations
 were carried out with B3LYP orbitals for 16 and 20 atoms in the supercell.
The cohesive energy obtained by the DMC method shows
an excellent agreement with experiment (Tab. II).
To evaluate the impact of the correlation on the gap we have
estimated the energy of the $ \Gamma$ $\to$ ${\rm B}$ excitation
 by an exciton calculation \cite{Lubos3}.
The QMC result for excitation energy is less perfect
with the difference from experiment being
a fraction of an eV. This clearly shows that the wavefunction
and corresponding fixed-node error is larger for an excited
state.
\begin{figure}[ht]
\hspace{-0.4cm}
\includegraphics[width=8cm]{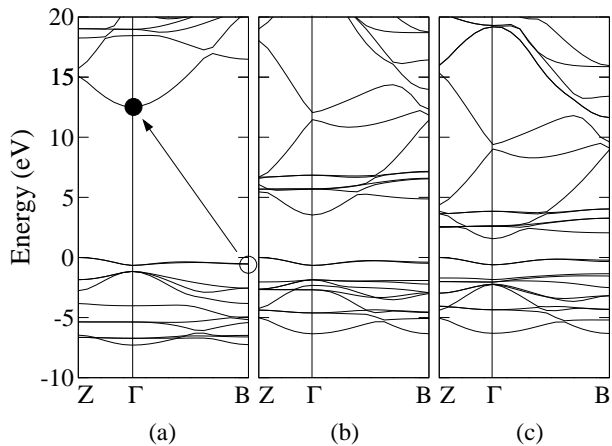}
\caption{The band structures of MnO obtained by (a) the unrestricted
 Hartree-Fock method, (b) DFT/B3LYP functional, and (c) DFT/PW86
 functional. The PW91 and PBE functionals provide essentially the
same picture as the PW86 functional.
 The calculated excitation in QMC is indicated by an arrow on the
UHF plot and the corresponding one-particle states
are denoted by open and filled circles.}
\end{figure}
Nevertheless, the  
 differential energy gain for  excited  vs. ground state
 from correlation of $\approx$
 8 eV is substantial and 
  demonstrates
 the importance of this effect both in qualitative
 and quantitative sense.
Due to large computational demands, 
 similar but statistically less precise
 DMC calculations were carried out also with the UHF orbitals. 
 While for the ground state the difference between the two sets of orbitals
 was marginal, the excited state with UHF orbitals
 appeared higher in energy approximately by 
 $\approx 1.5 (0.5)$ eV indicating thus, not surprisingly, 
 even larger fixed-node bias for the excited state in the UHF approach.

\begin{table}[ht]
\vspace{0.05cm}
\caption{The MnO solid cohesive energy and ${\rm B} \to \Gamma $ excitation energy
calculated by UHF, DFT and DMC methods compared with experiment.
The determinantal part of the DMC wavefunction used
the B3LYP one-particle orbitals.}
\begin{center}
\begin{tabular}{l c c c c c}
 & {UHF}
 & {PW86}
 & {B3LYP}
 & {DMC} 
 & {Exp.} \\
\hline
 E$_{\rm coh}$         & 6.03  & 11.00  & 9.21     & 9.40 (5) & 9.50 \\
${\rm B} \to \Gamma $  & 13.5$\;\;$ & 1.2  & 4.0     & 4.8 (2)  & $\approx$ 4.1 \\
\hline
\end{tabular}
\end{center}
\label{tab_yy}
\end{table}

It is quite encouraging that  the fixed-node DMC with
the simplest possible single-determinant
wavefunction leads to a consistent and parameter-free description 
of the basic properties
of this strongly correlated system.
 An obvious question to ask is whether one-determinant
 is sufficiently accurate for an antiferromagnet since the
 wavefunction with different spin-up and spin-down orbitals is 
 manifestly not
 an eigenfunction of the square of the total spin operator.
 In order to
  eliminate the spin contamination one would need to explore
 wavefunction forms beyond the single-determinant Slater-Jastrow, 
 for example,  generalized valence bond wavefunctions.
 However, since the actual mechanism is the Mn-O-Mn superexchange,
 one can expect the resulting effect to be small
 and most probably undetectable within our error bars (our calculations 
 with wavefunctions beyond the single
 UHF determinant were not conclusive.)

\begin{figure}
\hspace{-0.4cm}
\includegraphics[width=8cm]{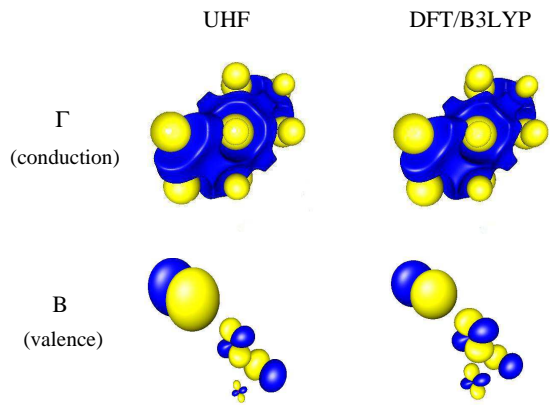}
\caption{Isosurfaces of UHF and DFT/B3LYP (blue/dark region is 
positive while yellow/light is negative). 
The valence B states are plotted for four atoms while the 
$\Gamma$ states are plotted for a supercell so that their conducting
character is visible.
The valence states  have
significant weights both from Mn $d$ states and O $p$ states.
Note small increase/decrease of $d/p$ orbital 
amplitudes from UHF to DFT/B3LYP.
The conduction state is much more delocalized and 
is composed from Mn $4s$ and O $3s$ atomic states.
}
\end{figure}

It is interesting to revisit now the one-particle results
and provide some feedback from our QMC calculations.
The analysis of orbitals indicates that
the nature of the top valence bands is rather similar in
all approaches with
both $p$ and $d$ states having 
significant weights in these states across the Brillouin zone
(see Fig. 2). This 
is clear also from the Mullikan population analysis which shows 
effective magnetic moments on Mn atoms  
in UHF, B3LYP and PW86/PW91/PBE methods to be 
4.92, 4.84 and
4.78$\mu_B$, respectively; these values are quite close
to each other and border the
range of experimental estimates of 4.58-4.78 $\mu_B$.

 The bottom of the conduction band is free-electron-like 
 $\Gamma$ state with significant amplitudes
from atomic O(3s) and Mn(4s) orbitals and it is this 
state which is responsible for the DFT gap closing  
in ferromagnetic or spin-disordered phases. 
For the ferromagnetic phase
 B3LYP exhibits a gap of $\approx$ 2.4 eV
and it is straightforward to check that by decreasing
the weight of exact exchange the gap decreases.
For example, with  10\% of exact exchange
in B3LYP the gap lowers to $\approx$ 1.2 eV. 
 The functionals without the exact exchange, such as 
PW86/PW91/PBE, lead to ferromagnetic metals
due to the overlaps with the uppermost valence bands
and subsequent 
rehybridization of states around the Fermi level. 
This
is the ``usual" DFT band gap problem which favors smooth and extended
states at the expense of the more localized ones which are  
stabilized by the exact (nonlocal) exchange and correlation effects. 

 The results presented here therefore
provide quite suggestive insights into the
 problem of
 a simple one-particle model appropriate for TMOs.
 In the Mott-Hubbard picture the origin of the band gap is
 the large on-site repulsion between the $d$ electrons
 which basically relates the gap to the two-site
 $d^nd^n \to
  d^{n-1}d^{n+1} $ type of excitation. 
 On the other hand, the charge
 transfer favors the ionic picture of Mn$^{++}$O$^{--}$
 with the oxygen $p$ states at the top of the valence band
 the gap given by the $p\to d$ excitation energy.
 Our QMC results  quantify that UHF produces 
 qualitatively correct wavefunctions 
 although biased towards the charge transfer limit, especially
for excited states.
 The hybrid B3LYP functional provides more balanced
  zero-order theory, arguably better
 and less biased than DFT non-hybrid functionals.  This view
 is supported also  by the orbital analysis proposed by Brandow
\cite{Brandowold,Brandownew},
 and, in effect, also by introduction of on-site terms which restore
 some of the Hartree-Fock character. 
To a certain extent, the hybrid functional alleviates the DFT 
 biases
 by eliminating part of the self-interaction and by
introducing the exchange ``pull-down" attraction resulting in lower 
energies of
localized states. Using such orbitals in QMC correlated framework 
enables us to obtain results which are close to experiment without 
any additional parameters.

It is tempting to consider the usefulness of
 a hybrid functional which  would be,
however, derived from a fundamental theory  
instead of a fit to a testing set of molecules underlying B3LYP 
{\cite{Becke}}. Although such approach would not fix all the 
deficiencies of the approximate DFT functionals it could serve, for example,
as a cost effective method for providing
 more appropriate sets of one-particle orbitals for building 
accurate wavefunctions. In fact, 
for molecular systems such as TiO and MnO we were able to directly optimize 
the weight of the exact exchange within a QMC framework.
By varying 
the weight of the exact exchange in
the functional and by iterating fixed-node DMC calculations
 we found the best set of one-particle orbitals which provided
the lowest fixed-node energy in DMC {\cite {Lucas}}; 
for solids such calculations might be possible in the near future.

 In conclusion, we have carried out calculations of the MnO solid
 in the variational and diffusion Monte Carlo methods. We have evaluated
 the cohesive and excitation energies which show excellent agreement with
 experiment.  
 We have pointed out the
 necessity of using high accuracy Ne-core pseudopotentials for Mn because
 of semicore $3s$ and $3p$ states. 
 The results  clearly
 show a crucial role of both exchange and correlation and their
 accurate description not only for MnO but obviously for
 other transition metal oxide systems as well. 

This work was supported by the ONR/DARPA grants ONR-N00014-01-1-1062
and ONR-N00014-01-1-0408.
Calculations were done at
 is done on the TITAN-IA64 cluster at NCSA, 
at PSC and at PLUTO-P4 cluster in KISTI (Korea 
Institute of Science and Technology Information). L.M. would like to thank
Yoonseok Lee for generously providing his pseudopotentials 
for testing.

\vfill
\end{document}